\documentclass[showpacs,amsmath,amssymb,prd,floatfix,twocolumn]{revtex4}

\usepackage{epsfig}
\usepackage{graphicx}
\usepackage{bm}
\usepackage{amsfonts}

\def\Ol{\Omega_{DE}}
\def\Olo{\Omega_{DE_0}}

\def\p{\partial}

\newcommand{\s}{\sigma}
\newcommand{\la}{\lambda}
\newcommand{\vp}{V_\phi(\phi)}
\newcommand{\vs}{V_\sigma(\sigma)}

\begin{document}

\title{Quintom dark energy models with nearly flat potentials}

\author{M. R. Setare}
 \affiliation{Department of Science, Payame Noor University,\\
Bijar, Iran }

\author{E. N. Saridakis }
 \affiliation{Department of Physics,
University of Athens,\\ GR-15771 Athens, Greece}

\begin{abstract}
We examine quintom dark energy models, produced by the combined
consideration of a canonical and a phantom field, with nearly flat
potentials and dark energy equation-of-state parameter $w_{DE}$
close to $-1$. We find that all such models converge to a single
expression for $w_{DE}(z)$, depending only on the initial field
values and their derivatives. We show that this quintom paradigm
allows for a description of the transition through $-1$ in the
near cosmological past. In addition, we provide the necessary
conditions for the determination of the direction of the
$-1$-crossing.
\end{abstract}

 \maketitle

\section{Introduction}

It is strongly believed nowadays that the universe is experiencing
an accelerating expansion, and this is supported by many
cosmological observations, such as SNe Ia \cite{1}, WMAP \cite{2},
SDSS \cite{3} and X-ray \cite{4}. These observations suggest that
the universe is dominated by dark energy, which provides the
dynamical mechanism for the accelerating expansion of the
universe. Furthermore, they suggest that the dark energy
equation-of-state parameter $w_{DE}$ might have crossed the
phantom divide $-1$ \cite{c14} from above in the near past.

In order to provide theoretical paradigms for the description of
dark energy, one can either consider theories of modified gravity
\cite{ordishov}, or field models of dark energy. The field models
that have been discussed widely in the literature consider a
cosmological constant \cite{cosmo}, a canonical scalar field
(quintessence) \cite{quint}, a phantom field, that is a scalar
field with a negative sign of the kinetic term
\cite{phant,phantBigRip}, or the combination of quintessence and
phantom in a unified model named quintom \cite{quintom}. The
advantage of quintom models is that they are capable of describing
the crossing of the phantom divide, since in quintessence and
phantom models $w_{DE}$ can indeed vary, but it always remains on
the same side of the phantom bound ($w_{DE}>-1$ in quintessence
and $w_{DE}<-1$ in phantom scenarios).

On the other hand, dark energy models with nearly flat potentials
have been shown to present interesting cosmological behavior,
especially in the case where $w_{DE}$ is around $-1$. Although
there is not a concrete proof, there are many arguments indicating
that nearly flat potentials are a natural way of acquiring
$w_{DE}\approx-1$, if one desires to avoid sophisticated, and
difficult to be justified, constructions. In the case of
quintessence and thawing quintessence the corresponding conclusion
has been acquired in various ways \cite{quinteflat}, while in the
case of phantom cosmological paradigm in \cite{Kujat}. Finally,
note that flat potentials, which keep the variations of the scalar
fields from their initial to their present values small, can also
be efficient to avoid unknown quantum gravity effects
\cite{Huang}.

In \cite{Scherrer:2007pu} the authors showed that all quintessence
models with nearly flat potentials converge to a single function
for $w(a)$, with $a$ the scale factor, which can be approximately
given analytically. Similarly, in \cite{Scherphan} the authors
result to such a limiting behavior for the phantom scenario with
nearly flat potentials. Both works share the general feature of
quintessence and phantom paradigms, that is they cannot describe
the $-1$ crossing, remaining either above (quintessence) or below
(phantom).

In the present work we are interested in investigating the
behavior of the combined case, that is we study quintom models
with nearly flat potentials. Note that we do not assume complete
dark energy domination, since matter is always non-negligible. We
provide analytically an approximated universal behavior for
$w_{DE}(z)$, which can naturally describe the crossing of the
phantom divide from above in the near past. This feature could
bring the model at hand closer to observations.

In section \ref{model} we construct general quintom models with
nearly flat potentials, and we extract the expression for
$w_{DE}(z)$. In section \ref{cosmimpl} we examine its behavior and
we discuss the cosmological implications. Finally, in section
\ref{conclusions} we summarize our results.

\section{Quintom Models with Nearly Flat Potentials}\label{model}

Throughout the work we consider a flat Robertson-Walker metric:
\begin{equation}\label{metric}
ds^{2}=dt^{2}-a^{2}(t)d\textbf{x}^2,
\end{equation}
with $a$ the scale factor. The action of a universe constituted of
a canonical $\phi$ and a phantom $\s$ fields is \cite{quintom}:
\begin{eqnarray}
S=\int d^{4}x \sqrt{-g} \left[\frac{1}{2} R
-\frac{1}{2}g^{\mu\nu}\p_{\mu}\phi\p_{\nu}\phi+\vp+\right.\nonumber\\
\left.+\frac{1}{2}g^{\mu\nu}\p_{\mu}\sigma\p_{\nu}\sigma+\vs
+\cal{L}_\text{M}\right], \label{actionquint}
\end{eqnarray}
where we have set $8\pi G=1$. The term $\cal{L}_\text{M}$ accounts
for the matter content of the universe, which for simplicity is
considered us dust. The Friedmann equations and the evolution
equation for the canonical and phantom fields are \cite{quintom}:
\begin{equation}
H^{2}=\frac{1}{3}\left[\rho_{M}+\rho_{\phi}+\rho_{\s}\right],
\label{Fr1}
\end{equation}
\begin{equation}
\left(\frac{\ddot{a}}{a}\right)=-\frac{1}{3}\left[\frac{\rho_M}{2}+2
p_{\phi}+2 p_{\s}+V_\phi(\phi)+V_\s(\s)\right], \label{Fr2}
\end{equation}
\begin{equation}
\ddot{\phi}+3H\dot{\phi}+\frac{\partial V_\phi(\phi)}{\partial
\phi}=0, \label{canonical}
\end{equation}
\begin{equation}
\ddot{\s}+3H\dot{\s}-\frac{\partial V_\s(\s)}{\partial \s}=0,
\label{phantom}
\end{equation}
where $H=\dot{a}/a$ is the Hubble parameter. In these expressions,
$p_\phi$ and $\rho_{\phi}$ are respectively the pressure and
density of the canonical field, while  $p_\s$ and $\rho_{\s}$ are
the corresponding quantities for the phantom field. Finally,
$\rho_M$ is the density of the matter content of the universe.

The energy density and pressure of the canonical and the phantom
fields, are given by:
\begin{eqnarray}
\rho_{\phi}=\frac{1}{2}\dot{\phi}^{2}+V_\phi(\phi)\nonumber\\
p_{\phi}=\frac{1}{2}\dot{\phi}^{2}-V_\phi(\phi)\label{enercanon}
\end{eqnarray}
and
\begin{eqnarray}
\rho_{\s}=-\frac{1}{2}\dot{\s}^{2}+V_\s(\s)\nonumber\\
p_{\s}=-\frac{1}{2}\dot{\s}^{2}-V_\s(\s).\label{enerphantom}
\end{eqnarray}
As usual, the dark energy of the universe is attributed to the
scalar fields and it reads:
\begin{equation}
\Ol=\frac{1}{3H^2}\left[\frac{1}{2}\left(\dot{\phi}^{2}-\dot{\s}^{2}\right)+V_\phi(\phi)+V_\s(\s)\right]
\label{OmL}.
\end{equation}
Finally, the equation of state for the  quintom dark energy is
\cite{quintom}:
\begin{equation}
w_{DE} =\frac{p_\phi+p_\s}{\rho_\phi+\rho_\s}=
\frac{\dot{\phi}^{2}-\dot{\s}^{2}-2[V_\phi(\phi)+V_\s(\s)]}
{\dot{\phi}^{2}-\dot{\s}^{2}+2[V_\phi(\phi)+V_\s(\s)]}\label{eqstate}.
\end{equation}

Using the definitions for the energy densities and pressures
(\ref{enercanon}),(\ref{enerphantom}), the Friedmann equations
(\ref{Fr1}),(\ref{Fr2}) can be re-written as:
\begin{equation}
\label{sys1}
\dot{H}=-\frac{1}{2}\left(\rho_M+\dot{\phi}^2-\dot{\s}^2 \right)
\end{equation}
\begin{equation}\label{sys2}
H^2=\frac{1}{3}\left[\rho_M+\frac{1}{2}\dot{\phi}^2+V_\phi(\phi)-
\frac{1}{2}\dot{\s}^2+V_\s(\s)\right].
\end{equation}
Lastly, the equations close by considering the evolution of the
matter density:
\begin{equation}\label{sys3}
\dot{\rho}_M=-3H\rho_M.
\end{equation}

In order to provide an analytical expression for $w_{DE}$ we have
to transform the dynamical system
(\ref{canonical}),(\ref{phantom}),(\ref{sys1}),(\ref{sys2}) into
an autonomous form \cite{{Copeland}}. This will be achieved by
introducing the auxiliary variables:
\begin{equation}
x=\frac{\sqrt{\dot{\phi}^2-\dot{\s}^2}}{\sqrt{6}H}\label{auxilliary1}
  \end{equation}
\begin{equation}
y=\frac{\sqrt{V_\phi(\phi)+V_\s(\s)}}{\sqrt{3}H}\label{auxilliary2}
  \end{equation}
\begin{equation}
\la=-\frac{1}{V_\phi(\phi)+V_\s(\s)}\,
\frac{\dot{\phi}\frac{dV_\phi(\phi)}{d\phi}-\dot{\s}\frac{dV_\s(\s)}{d\s}}{\sqrt{\dot{\phi}^2-\dot{\s}^2}}\label{auxilliary3}
\end{equation}
\begin{equation}
\delta=-\frac{1}{V_\phi(\phi)+V_\s(\s)}\,\frac{\dot{\phi}\frac{dV_\phi(\phi)}{d\phi}+\dot{\s}\frac{dV_\s(\s)}{d\s}}{\sqrt{\dot{\phi}^2-\dot{\s}^2}}
\label{auxilliary4},
\end{equation}
together with $M=\ln a$. Thus, it is easy to see that for every
quantity $F$ we acquire $\dot{F}=H\frac{dF}{dM}$.

Using these variables, we result to the following system:
\begin{equation}
\frac{dx}{dM}=\frac{3}{2}x \left(1+x ^2-y ^2\right)- 3x
+\sqrt{\frac{3}{2}}\la \, y ^2\label{autonomous1}
\end{equation}
\begin{equation}
\frac{dy }{dM}=\frac{3}{2}y  \left(1+x ^2-y ^2\right)-
\sqrt{\frac{3}{2}}\delta \,x  y \label{autonomous2}
\end{equation}
\begin{eqnarray}
&\frac{d\la}{dM}=\frac{A_\la(y,x\la,x\delta)}{V_\phi(\phi)+V_\s(\s)}\,\frac{dV_\phi(\phi)}{d\phi}+\frac{B_\la(y,x\la,x\delta)}{V_\phi(\phi)+V_\s(\s)}\,\frac{d^2V_\phi(\phi)}{d\phi^2}+\nonumber\\
&\frac{C_\la(y,x\la,x\delta)}{V_\phi(\phi)+V_\s(\s)}\,\frac{dV_\s(\s)}{d\s}+\frac{D_\la(y,x\la,x\delta)}{V_\phi(\phi)+V_\s(\s)}\,\frac{d^2V_\s(\s)}{d\s^2}\label{autonomous3}
\end{eqnarray}
\begin{eqnarray}
&\frac{d\delta}{dM}=\frac{A_\delta(y,x\la,x\delta)}{V_\phi(\phi)+V_\s(\s)}\,\frac{dV_\phi(\phi)}{d\phi}+\frac{B_\delta(y,x\la,x\delta)}{V_\phi(\phi)+V_\s(\s)}\,\frac{d^2V_\phi(\phi)}{d\phi^2}+\nonumber\\
&\frac{C_\delta(y,x\la,x\delta)}{V_\phi(\phi)+V_\s(\s)}\,\frac{dV_\s(\s)}{d\s}+\frac{D_\delta(y,x\la,x\delta)}{V_\phi(\phi)+V_\s(\s)}\,\frac{d^2V_\s(\s)}{d\s^2}\label{autonomous4}.
\end{eqnarray}
The complicated functions
$A_\la,B_\la,C_\la,D_\la,$$A_\delta,B_\delta,C_\delta,D_\delta$
can be straightforwardly calculated through differentiation of
(\ref{auxilliary3}) and (\ref{auxilliary4}), and elimination of
$\dot{\phi}$ and $\dot{\s}$ in terms of $x,y,\la,\delta$. However,
their exact form is not needed for the purpose of this work.

In terms of these auxiliary variables $\Ol$ from relation
(\ref{OmL}) can be written as:
\begin{equation}
\Omega_{DE} =x^2+y ^2.
 \label{omega}
\end{equation}
Similarly, using  (\ref{eqstate}) we can find the corresponding
relation for $w_{DE}$. However, it is more convenient to define a
new variable $\zeta=1+w_{DE}$, which simply reads:
\begin{equation}\label{eqstate2}
 \zeta=1+w_{DE}=\frac{2x ^2}{x^2+y^2}.
\end{equation}
Thus, we can easily see that:
\begin{eqnarray}
 &&x^2=\frac{\zeta\Omega_{DE}}{2} \nonumber\\
&&y^2=\Omega_{DE}\left(1-\frac{\zeta}{2}\right)\label{relations}.
\end{eqnarray}
Therefore, by differentiating these expressions with respect to
$M=\ln a$ and using (\ref{autonomous1}),(\ref{autonomous2}), we
acquire the autonomous equations for $\Ol$ and $\zeta$: {\small{
\begin{equation}
\frac{d\Omega_{DE}}{dM}=3\Omega_{DE}(1-\zeta)(1-\Omega_{DE})+\Omega_{DE}\sqrt{3\zeta\Omega_{DE}}\left(1-\frac{\zeta}{2}\right)\,\theta\label{relations2O}
 \end{equation}}}
 \begin{eqnarray}
 \frac{d\zeta}{dM}=-3\zeta(2-\zeta)+\la(2-\zeta)\sqrt{3\zeta\Omega_{DE}}-\nonumber\\
 -\sqrt{3\zeta\Omega_{DE}}\left(1-\frac{\zeta}{2}\right)\zeta\,\theta\label{relations2z},
\end{eqnarray}
where
\begin{equation}\label{theta}
 \theta=\la-\delta.
\end{equation}
Finally, we obtain:
\begin{eqnarray} \label{diffeqn}
&\frac{d\zeta}{d\Omega_{DE}}=\frac{\frac{d\zeta}{dM}}{\frac{d\Omega_{DE}}{dM}}=\nonumber\\
&=\frac{-3\zeta(2-\zeta)+\la(2-\zeta)\sqrt{3\zeta\Omega_{DE}}-\sqrt{3\zeta\Omega_{DE}}\left(1-\frac{\zeta}{2}\right)\zeta\,\theta}
{3\Omega_{DE}(1-\zeta)(1-\Omega_{DE})+\Omega_{DE}\sqrt{3\zeta\Omega_{DE}}\left(1-\frac{\zeta}{2}\right)\,\theta}.
\end{eqnarray}

Equation (\ref{diffeqn}), which gives the autonomous evolution for
$\zeta$ (i.e for the equation-of-state parameter) is exact and
takes into account the full dynamics of the system evolution,
which is also determined exactly by equations
(\ref{autonomous1})-(\ref{autonomous4}). In order to proceed to
the extraction of analytical solutions we have to make two
assumptions. This will allow us to bypass the details of specific
models and describe the general behavior of quintom scenarios with
nearly flat potentials. Fortunately, as we are going to see in the
next section, the error of our approximated solution comparing to
the numerical elaboration of the exact system is small.

The first approximation is that $|\zeta|\ll1$, i.e $w_{DE}$ is
close to $-1$. This assumption is justified by observations both
at present time and in the recent cosmological past. The second
assumption is that the potentials of the model are nearly flat,
which is the case of interest of the present work. In particular
we assume:
\begin{eqnarray}
&&\left|\frac{1}{V_\phi(\phi)+V_\s(\s)}\,\frac{dV_\phi(\phi)}{d\phi}\right|\ll1\nonumber\\
  &&\left|\frac{1}{V_\phi(\phi)+V_\s(\s)}\,\frac{dV_\s(\s)}{d\s}\right|\ll1\nonumber\\
   &&\left|\frac{1}{V_\phi(\phi)+V_\s(\s)}\,\frac{d^2V_\phi(\phi)}{d\phi^2}\right|\ll1\nonumber\\
  &&\left|\frac{1}{V_\phi(\phi)+V_\s(\s)}\,\frac{d^2V_\s(\s)}{d\s^2}\right|\ll1
 \label{assumption2}.
\end{eqnarray}
Therefore, the variables $\lambda$ and $\delta$ can be assumed to
be constant, and equal to their initial values: {\small{
\begin{equation}
\la=\la_0=\left.-\frac{1}{V_\phi(\phi)+V_\s(\s)}\,
\frac{\dot{\phi}_0\frac{dV_\phi(\phi)}{d\phi}-\dot{\s}_0\frac{dV_\s(\s)}{d\s}}{\sqrt{\dot{\phi}_0^2-\dot{\s}_0^2}}\right|_{\phi=\phi_0,\sigma=\sigma_0}
\label{la0}
\end{equation}
\begin{equation}
\delta=\delta_0=\left.-\frac{1}{V_\phi(\phi)+V_\s(\s)}\,\frac{\dot{\phi}_0\frac{dV_\phi(\phi)}{d\phi}+\dot{\s}_0\frac{dV_\s(\s)}{d\s}}
{\sqrt{\dot{\phi}_0^2-\dot{\s}_0^2}}\right|_{\phi=\phi_0,\sigma=\sigma_0}
\label{delta0},
\end{equation}}}
with $\phi_0$, $\s_0$ the initial  values and $\dot{\phi}_0$,
$\dot{\s}_0$ the initial derivatives of the fields just before
they begin the roll-down to the potential (equivalently the
corresponding values in the near cosmological past). Thus,
$|\la_0|,|\delta_0|\ll1$, too. Note that at first sight, one could
think that this approximation would brake down if $\dot{\phi}_0$
and $\dot{\s}_0$ are equal, or more generally if  $\dot{\phi}$ and
 $\dot{\s}$ become equal at any time. However, this is not
 happening since in this case $x$ in (\ref{auxilliary1}) and $\zeta$ in (\ref{eqstate2}) also go to zero,
with the limits $\la_0\sqrt{\zeta}$, $\delta_0\sqrt{\zeta}$ not
only regular but proportional to the first potential derivatives.
That is, equation (\ref{diffeqn}) becomes even simpler. Finally,
we mention that in the case where $\dot{\phi}$ and
 $\dot{\s}$ become equal at some time, the derivatives $\frac{d\la}{dM}$
 and$\frac{d\delta}{dM}$ remain small since they also depend on
 the combinations $x\la$ and $x\delta$.

Keeping terms up to lowest order in $\zeta$, $\lambda_0$,
$\delta_0$, (\ref{diffeqn}) yields:
\begin{equation}\label{diffeqn2}
\frac{d\zeta}{d\Omega_{DE}}=
\frac{-6\zeta+2\la_0\sqrt{3\zeta\Omega_{DE}}-\sqrt{3\zeta\Omega_{DE}}\zeta\,\theta_0}
{3\Omega_{DE}(1-\Omega_{DE})},
\end{equation}
where $\theta_0\equiv\la_0-\delta_0$. Equation (\ref{diffeqn2})
can be transformed into a linear differential equation under the
transformation $s=\sqrt{\zeta}$ and  can be solved exactly. The
solution for $\zeta=1+w_{DE}$ is:{\small{
\begin{eqnarray}
&1+w_{DE}(\Ol) =\left\{\frac{\sqrt{3}}{\theta_0\sqrt{\Ol}}+
\right.\nonumber\\
&\left.+\sqrt{\frac{3+2\theta_0\lambda_0}{\theta_0^2}} \
\tanh\left[\frac{\theta_0}{2\sqrt{3}}\sqrt{\frac{3+2\theta_0\lambda_0}{\theta_0^2}}
\,\ln \left(\frac{1-\sqrt{\Ol}} {1+\sqrt{\Ol}}
\right)\right]\right\}^2\label{wlfin}
\end{eqnarray}}}
Equation (\ref{wlfin}), along with the corresponding result for
$w(z)$ derived below, is our main result.  It shows that for
sufficiently flat potentials, all quintom models with
$w_{DE}\approx-1$ at present time, approach a single generic
behavior. The specific role of the potentials is to determine the
small constants $\lambda_0$ and $\theta_0$. Finally, we mention
that (\ref{wlfin}) gives always real values for $w_{DE}$ as
expected, due to the specific combination of the two terms that
may be imaginary.

In order to express our result in a form more suitable for
comparison with observations we will use (\ref{relations2O}) up to
lowest order in $\zeta$ and $\theta_0$, in order to acquire
$\Ol(M)$, i.e $\Ol(z)$, since $e^M\equiv a=(1+z)^{-1}$  (we set
the present value of the scale factor $a_0=1$). In this
approximation the solution is:
\begin{equation}
\label{Omz} \Ol(z) = \left[1 + \left(\Olo^{-1} - 1 \right)(1+z)^3
\right]^{-1},
\end{equation}
where $\Olo$ is the present-day value of $\Ol$. Note that equation
(\ref{Omz}) coincides with the expression for $\Ol$ in
\cite{Crittenden}, where the authors obtain in a different
framework but under the assumption $w_{DE}\approx-1$.

Substituting (\ref{Omz}) into (\ref{wlfin}) we obtain:
\begin{eqnarray}
1+w_{DE}(z)
=\left\{\frac{\sqrt{3}}{\theta_0}\left[1+(1+z)^3\left(\frac{1}{\Olo}-1\right)\right]^{1/2}+\right.\nonumber\\
\left.+
 \sqrt{\frac{3+2\theta_0\lambda_0}{\theta_0^2}}\,
\tanh\left\{\frac{\theta_0}{2\sqrt{3}}\sqrt{\frac{3+2\theta_0\lambda_0}{\theta_0^2}}
\right.\right.\nonumber\\
\left.\left.\ln
\left[\frac{\left[1+(1+z)^3\left(\frac{1}{\Olo}-1\right)\right]^{1/2}-1}
{\left[1+(1+z)^3\left(\frac{1}{\Olo}-1\right)\right]^{1/2}+1}
\right]\right\}\right\}^2\label{wlfin2}.
\end{eqnarray}

We can expand (\ref{wlfin2}) with $a=(1+z)^{-1}$ around $a=1$, and
acquire a linear relation of the form
$w_{DE}(a)=w_{{DE_0}}+w_{DE_a}(1-a)$, in accordance with
Chevallier-Polarski-Linder parametrization \cite{Lindp}. In this
case the parameters $w_{{DE_0}}$ and $w_{DE_a}$ are independent
and are given as functions of $\la_0$, $\theta_0$ and $\Olo$. This
is an advantage comparing to the simple quintessence
\cite{Scherrer:2007pu} and simple phantom \cite{Scherphan} cases
where such a procedure leads to a dependence of  $w_{DE_a}$ on
$w_{{DE_0}}$, in contrast with the imposed parametrization. The
inclusion of two fields, and thus of additional degrees of
freedom, in the model at hand, restores the linear parametrization
in its correct form, that is with two independent parameters.
Performing the aforementioned expansion imposing
$\Olo=0.73\pm0.03$, with $|\lambda_0|\leq0.5$ and
$|\theta_0|\leq0.5$, we obtain $-1.08\leq w_{{DE_0}}\leq-0.92$ and
$-0.14\leq w_{{DE_a}}\leq0.14$. These limits are narrower than
those arising from observations \cite{1,2,3,4}, which was expected
since in our analysis we have assumed that $w_{DE}(a)$ is close to
$-1$ ($|\zeta|\ll1$).

Before closing this section let us make some comments on the
expressions (\ref{wlfin}) and (\ref{wlfin2}). If we desire to
obtain the simple canonical field, that is the case of
quintessence models, we have to set the quantities that are
relevant to the $\s$-field to zero. Thus, (\ref{la0}) and
(\ref{delta0}) imply $\la_0=\delta_0\in\mathbb{R}$ and therefore
$\theta_0=0$. In this case (\ref{wlfin}) gives:
\begin{equation}
1 + w_{DE} = \frac{\lambda_0^2}{3}\left[\frac{1}{\sqrt{\Ol}} -
\frac{1}{2}\left(\frac{1}{\Ol} - 1 \right) \ln
\left(\frac{1+\sqrt{\Ol}} {1-\sqrt{\Ol}}
\right)\right]^2,\label{quintess}
\end{equation}
which is just the result obtained in \cite{Scherrer:2007pu}. In
addition, in this case by inserting the present value $w_{{DE_0}}$
one can solve for $\la_0$, and substituting in (\ref{wlfin2}) he
can obtain $w_{DE}(z)$ with only  $w_{{DE_0}}$ and $\Olo$ as
parameters. Note that for the quintessence scenario, the variable
$x$ defined in (\ref{auxilliary1}) is always real and thus
(\ref{eqstate2}) implies that $w_{DE}$ is always larger than -1.
This is just what is expected for a quintessence model.

On the other hand, if we desire to obtain the simple phantom
model, then we have to set the quantities relevant to the
$\phi$-field to zero. Therefore, $\delta_0=-\la_0\in\mathbb{I}$
and thus $\theta_0=2\la\in\mathbb{I}$. In this case one finds
exactly the same expression (\ref{quintess}), with $\la_0^2$ being
negative, which is just the result obtained in \cite{Scherphan}.
Finally, in this simple phantom case, one can also eliminate
$\la_0$ and acquire $w_{DE}(z)$ in terms of  $w_{{DE_0}}$ and
$\Olo$. We mention that now $x$ is purely imaginary (see relation
(\ref{auxilliary1})) and thus (\ref{eqstate2}) leads to $w_{DE}$
always smaller than $-1$. Again, this is what is expected for a
phantom model.

In the quintom model at hand, that is in the case where both the
canonical and the phantom fields are present, $\la_0$, $\delta_0$
and $\theta_0$ are purely real or purely imaginary, depending of
which field is dominant, and the three variables belong to the
same set each time. We stress that $\la$, $\delta$ and $\theta$,
as well as $x$, are just suitable variables which allows us to
transform the system to its autonomous form, and are not related
to any observables. Therefore, their purely imaginary character is
just a statement of the dominance of the phantom field, that is it
acquires a robust and physical content. This becomes obvious by
the fact that all observables are real. Indeed, we can easily see
that $\Ol$ and $w_{DE}$ in (\ref{omega}), (\ref{eqstate2}),
(\ref{wlfin}) and (\ref{wlfin2}) are always real in the case of
purely imaginary $\la$, $\delta$, $\theta$ and $x$. The only
effect is that $w_{DE}$ is below the phantom divide. But such a
transition is exactly the motive of the present work.

Finally, we mention that when a crossing of the phantom divide
takes place, that is when $x$ crosses zero, our system remains
regular, and this was expected since even the naively acquired
singular behavior is not related with the initial cosmological
equations but only with the transformation we use to solve them
analytically (one could resemble the case at hand with the
distinction between
 true singularities and coordinate ones in general relativity). In particular, in
this case all the limits not only do exist but are much smaller
than one, and the autonomous equations become even simpler. This
can be also confirmed by the observation that when $\phi\approx\s$
the cosmological equations (\ref{Fr1})-(\ref{phantom}) become
significantly simpler, and thus the corresponding autonomous
system is simpler, too. Lastly, the regular behavior of our
solution procedure is also confirmed by the numerical elaboration
of the next section.

\section{Cosmological implications}
\label{cosmimpl}

\begin{figure}[ht]
\begin{center}
\mbox{\epsfig{figure=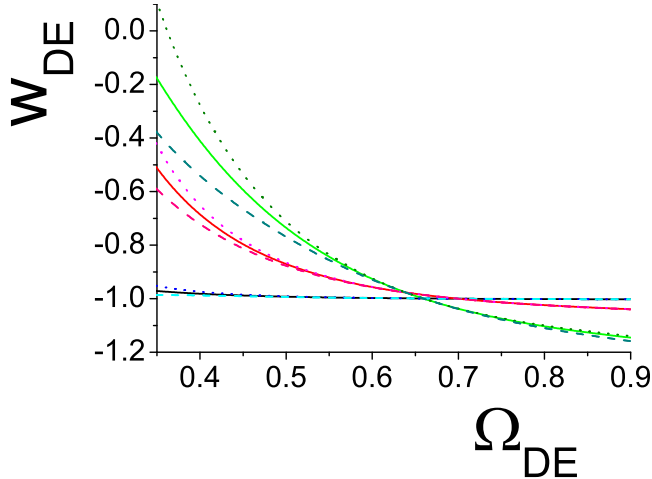,width=8.9cm,angle=0}}
\caption{(Color online) {\it  A comparison between the exact
result for $w_{DE}(\Ol)$ for $V_\phi(\phi)=\phi^2$,
$V_\s(\s)=\s^2$ (dotted curves),  the exact result for
$w_{DE}(\Ol)$ for $V_\phi(\phi)=\phi^{-2}$, $V_\s(\s)=\s^{-2}$
(dashed curves), and our approximated analytical result for
$w_{DE}(\Ol)$ provided by relation (\ref{wlfin}) (solid curves).
The group of curves from top to bottom correspond to  $\la_0=0.7
$, $\theta_0=0.7$, to $\la_0=0.4$, $\theta_0=0.4 $ and to
$\la_0=0.1$, $\theta_0=0.4$. The error of our analytical result
increases with the increase of $\la_0$ and $\theta_0$, and with
the increase of the distance of $w_{DE}$ from $-1$. }}
\label{wlomtest}
\end{center}
\end{figure}

Let us now investigate the cosmological implications of the
acquired results. First of all, we desire to check the accuracy of
our approximated analytical solution (\ref{wlfin}). In
fig.~\ref{wlomtest} we perform such a comparison. We have selected
two potential choices, namely $V_\phi(\phi)=\phi^2$,
$V_\s(\s)=\s^2$ (dotted curves) and $V_\phi(\phi)=\phi^{-2}$,
$V_\s(\s)=\s^{-2}$ (dashed curves) and we have numerically found
the exact $w_{DE}(\Ol)$ behavior of the cosmological system, for
three combinations of the parameters $\la_0$ and $\theta_0$
(fixing suitably the values of $\dot{\phi}_0$ and $\dot{\s}_0$).
From top to bottom the group of curves correspond to  $\la_0=0.7$,
$\theta_0=0.7$, to $\la_0=0.4$, $\theta_0=0.4$ and to $\la_0=0.1$,
$\theta_0=0.4$. In addition, for the same parameter values we have
used relation (\ref{wlfin}) to obtain the approximated analytical
behavior (solid curves). As can be seen, the deviation from the
exact solution is larger for larger distance of $w_{DE}$ from
$-1$, which was expected since we have approximated $|1+w_{DE}|$
to be much smaller than $1$. Secondly, we observe that the error
of our approximated result increases with the increase of the
parameters $\la_0$ and $\theta_0$, i.e from bottom to top, which
was also expected since we have assumed $|\la_0|,|\theta_0|\ll1$.
Thus, fig.~\ref{wlomtest} constraints the applicability range of
our approximated analytical solution to $-1.1<w_{DE}<-0.9$ and
$|\la_0|,|\theta_0|\lesssim0.5$. We mention that for higher
parameter values or for models with $w_{DE}$ larger than 0, the
deviation of our approximated solution from the exact cosmological
evolution can be dramatic, and our approximation scheme brakes
down. But these cosmological scenarios are beyond the purpose of
the present work. Finally, note that independently of the
subsequent cosmological evolution, negative power-law potentials
\cite{negpot} can fulfill the nearly-flat potential conditions
(\ref{assumption2}) if $\phi_0$ is sufficiently large. This
feature shows that any potential can give rise to the type of
models discussed here, as long us they satisfy
(\ref{assumption2}), and this was also shown in
\cite{Scherrer:2007pu} for the case of thawing quintessence
scenario.

\begin{figure}[ht]
\begin{center}
\mbox{\epsfig{figure=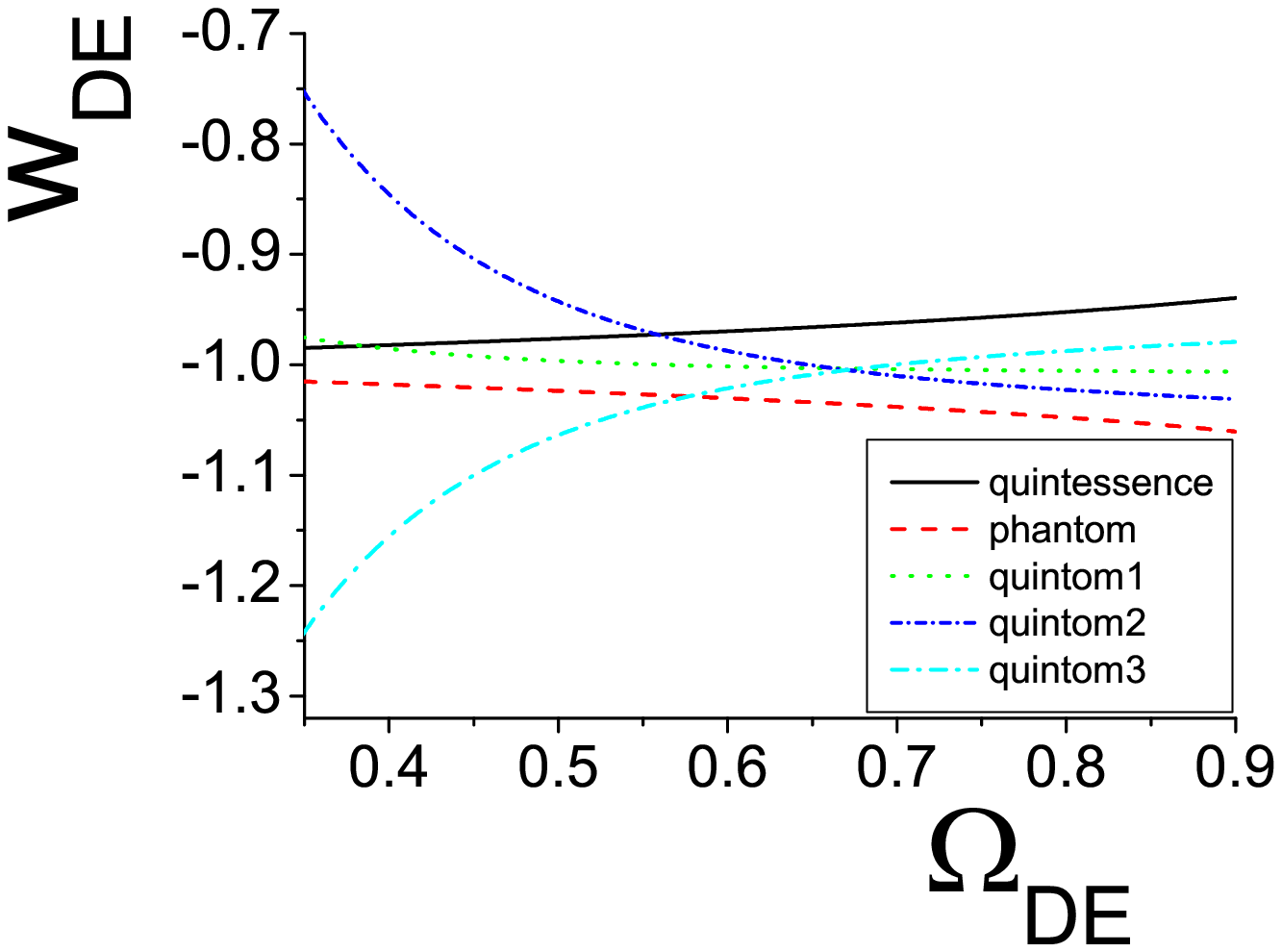,width=8.9cm,angle=0}}
\caption{(Color online) {\it  $w_{DE}(\Ol)$ provided by relation
(\ref{wlfin}) for various cosmological paradigms. The solid and
the dashed curves correspond to simple quintessence and simple
phantom models, with $\la_0^2=0.5$ and $\la_0^2=-0.5$
respectively. The dotted curve (quintom1) corresponds to
 $\la_0=0.1 $, $\theta_0=0.4 $. The sort-dashed-dotted curve
(quintom2) corresponds to  $\la_0=0.3 $, $\theta_0=0.1 $. The
dashed-dotted curve (quintom3) corresponds to $\la_0=0.3 i$,
$\theta_0=0.1 i$.}} \label{wlomega}
\end{center}
\end{figure}

Having determined the applicability area of our approximated
scheme we proceed to specific cosmological scenarios. In
fig.~\ref{wlomega} we depict $w_{DE}(\Ol)$, given by
(\ref{wlfin}), for five different models.
Firstly, with the solid and the dashed curves we present the
simple quintessence and the simple phantom scenarios, that is when
$\theta_0=0$ and $\la_0^2=0.5$,  $\la_0^2=-0.5$ respectively.
These results coincide with those of \cite{Scherrer:2007pu} and
\cite{Scherphan}, noting that in these works the authors draw
$w_{DE}(a)$ instead of $w_{DE}(z)$. As it was known, the simple
quintessence and the simple phantom models cannot describe the
transition through the phantom divide $-1$, and $w_{DE}$ remains
always on the same side of this bound ($w_{DE}>-1$ for
quintessence and $w_{DE}<-1$ for the phantom scenario). However,
the combined consideration of both models can indeed describe the
$-1$-crossing. In fig.~\ref{wlomega} we depict $w_{DE}(\Ol)$ for
three different quintom models. The dotted curve (quintom1)
corresponds to $\la_0=0.1 $, $\theta_0=0.4 $, the
sort-dashed-dotted curve (quintom2) corresponds to  $\la_0=0.3 $,
$\theta_0=0.1 $ and the dashed-dotted curve (quintom3) corresponds
to $\la_0=0.3 i$, $\theta_0=0.1 i$. As we observe, all three
models present the phantom-divide crossing.

\begin{figure}[ht]
\begin{center}
\mbox{\epsfig{figure=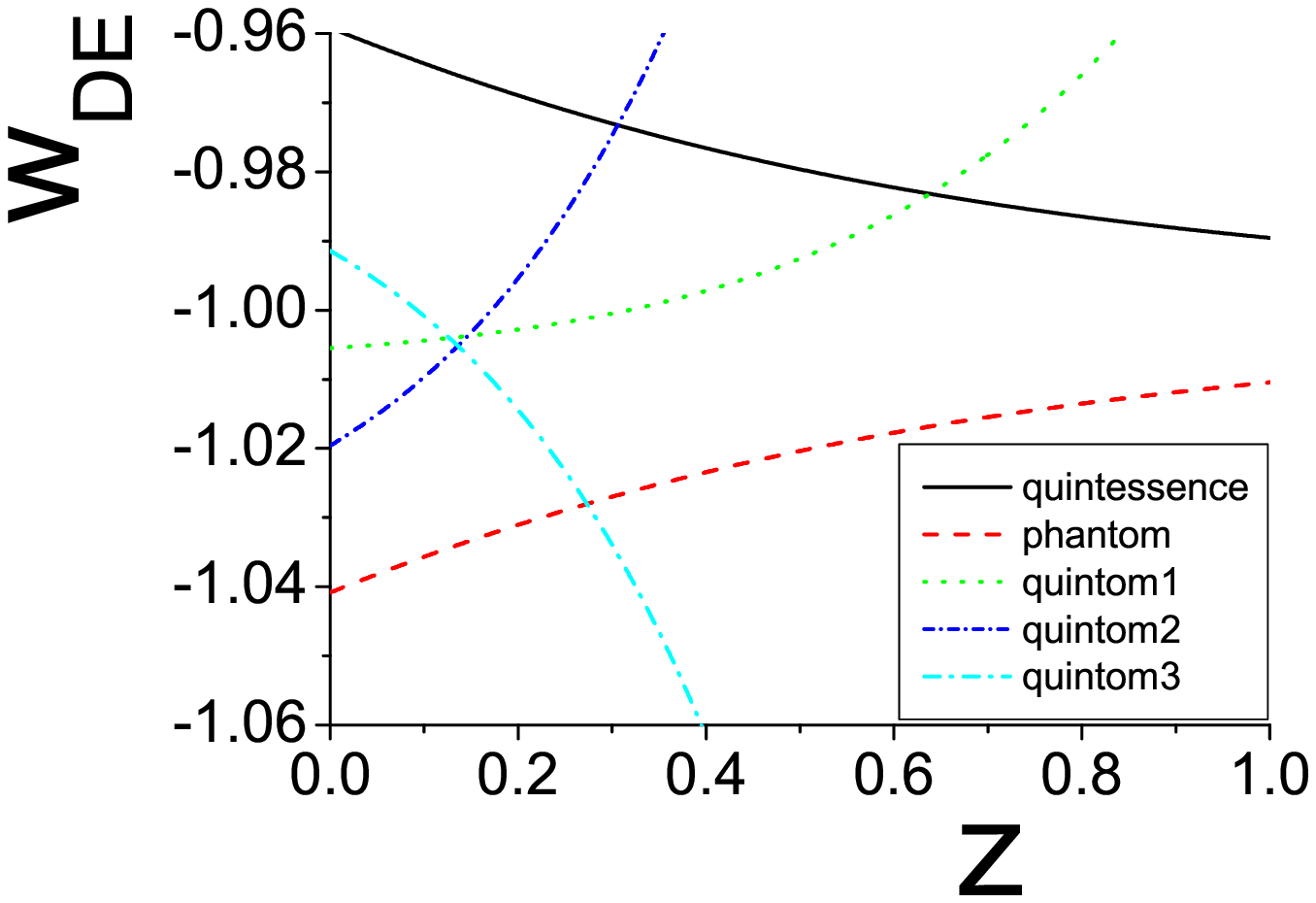,width=8.9cm,angle=0}}
\caption{(Color online) {\it $w_{DE}(z)$ provided by relation
(\ref{wlfin2}) with $\Olo\approx0.73$, for various cosmological
paradigms. The solid and the dashed curves correspond to simple
quintessence and simple phantom models, with $\la_0^2=0.5$ and
$\la_0^2=-0.5$ respectively. The dotted curve (quintom1)
corresponds to
 $\la_0=0.1 $, $\theta_0=0.4 $. The sort-dashed-dotted curve
(quintom2) corresponds to  $\la_0=0.3 $, $\theta_0=0.1 $. The
dashed-dotted curve (quintom3) corresponds to $\la_0=0.3 i$,
$\theta_0=0.1 i$. }} \label{wlzeta}
\end{center}
\end{figure}

In order to acquire a more transparent picture, in
fig.~\ref{wlzeta} we depict  $w_{DE}(z)$ (given by relation
(\ref{wlfin2}) with $\Olo\approx0.73$) for the same cosmological
models of fig.~\ref{wlomega}.
As we observe, although all three examined quintom models can
describe the phantom-divide crossing, only quirnom1 and quintom2
present the desired behavior, that is $w_{DE}(z)$ crossing $-1$
from above, resulting to $w_{{DE_0}}<-1$ at present, as it might
be the case according to observations. We remind that according to
(\ref{la0}) and (\ref{delta0}), imaginary values correspond to
initial phantom dominance, while real parameter values correspond
to initial canonical-field dominance. Thus our framework reveals a
way of determining the specific features of a quintom model. In
particular, the crossing through $-1$ requires a sign change of
$x^2$ (see relation (\ref{eqstate2})), and thus a change in which
field's kinetic term is larger (see definition
(\ref{auxilliary1})). If initially (with ``initial'' having the
meaning of the beginning of slow-roll as considered in
\cite{Scherrer:2007pu,Scherphan}) is the canonical field that is
dominant and at some point the phantom field dominates the
evolution, then the crossing of $-1$ is from above to below. This
case corresponds to quintom1 and quintom2 paradigms of
fig.~\ref{wlzeta}. On the other hand, if initially the evolution
is dominated by the phantom field, and progressively by the
canonical field, then the crossing through $-1$ takes place from
below to above. This is the case of quintom3 scenario of
fig.~\ref{wlzeta}. Therefore, in order to describe the possible
crossing of $w_{DE}(z)$ one has to consider an initial canonical
field dominance, with a subsequent increase and dominance of the
phantom field. These conditions are necessary for a certain
$-1$-crossing, bur are not efficient. That is, if the initial
dominance of one of the fields is sufficiently strong, then
$w_{DE}(z)$, although moving towards $-1$, it will never succeed
to cross it.

Let us close this section with a quantitative discussion. Current
observations of $w_{DE}(z)$ \cite{1,2,3,4} can accept the
description of quintom models with nearly flat potentials
presented here. The most significant advantage of this scenario is
the capability of describing the crossing through the phantom
divide from above to below, in the near cosmological past, if such
a crossing will be retained by future and more exact observations.
In addition, even without a crossing, the model at hand can
describe the $w_{DE}(z)$-evolution in close agreement with
current observational limits. However, we mention that such a
description is quantitatively trustworthy if the current value
$w_{{DE_0}}$ is larger than $-1.1$ and smaller than $-0.9$. On the
contrary, if $w_{{DE_0}}\lesssim-1.1$ or $w_{{DE_0}}\gtrsim-0.9$,
then the errors of our approximated analytical solution
(\ref{wlfin2}) become relatively large and the results have to be
considered only qualitatively.

\section{Conclusions}
\label{conclusions}

In this work we examine quintom models with nearly flat
potentials, without the assumption of complete dark energy
domination. In the case where the dark energy equation-of-state
parameter $w_{DE}$ is close to $-1$, we provide analytically an
approximated universal expression for $w_{DE}(z)$ for all such
models. This expression depends only on the initial conditions
(beginning of slow roll), i.e on the values of the potentials and
their derivatives at a specific point, and on the values of the
field kinetic terms at the same point.  This feature arises
because, due to the potential flatness, the fields never roll very
far along the potentials in order  to ``feel" the rest of their
shape.

Contrary to the case of simple quintessence or simple phantom
models, where $w_{DE}(z)$ is always on the same side of the
phantom divide ($w_{DE}>-1$ for quintessence and $w_{DE}<-1$ for
the phantom scenario) the quintom paradigm allows for a
description of the transition through $-1$. In addition, we
provide the necessary conditions for a specific such crossing. In
particular, if initially the universe is dominated by the
canonical field then the subsequent evolution can bring the
phantom field domination and thus the $-1$-crossing from above to
below. On the other hand, if initially is the phantom field that
dominates then the evolution can lead to $-1$-crossing from below
to above. Thus, a not-very-strong initial dominance of the
canonical field can lead to a cosmological evolution where
$w_{DE}(z)$  crosses $-1$ from above to below in the near past. In
conclusion, the determination of the impact of the two fields, can
provide a $w_{DE}(z)$-evolution in agreement with current
observational limits.\\

\paragraph*{{\bf{Acknowledgements:}}}
E. N. Saridakis wishes to thank Institut de Physique Th\'eorique,
CEA, for the hospitality during the preparation of the present
work.


\begin{thebibliography}{99}

\bibitem{1}
S. Perlmutter {\it{et al.}} [Supernova Cosmology Project
Collaboration], Astrophys. J. {\bf 517}, 565 (1999).

\bibitem{2}
C. L. Bennett {\it{et al.}}, Astrophys. J. Suppl. {\bf 148}, 1
(2003).

\bibitem{3}
M. Tegmark {\it{et al.}} [SDSS Collaboration], Phys. Rev. D {\bf
69}, 103501 (2004).

\bibitem{4}
S. W. Allen, {\it{et al.}}, Mon. Not. Roy. Astron. Soc. {\bf 353},
457 (2004).

\bibitem{c14}
 P. S. Apostolopoulos, and N. Tetradis, Phys. Rev. D {\bf
74}, 064021 (2006); H.-S. Zhang, and Z.-H. Zhu, Phys. Rev. D {\bf
75}, 023510 (2007).

\bibitem{ordishov}
S.Nojiri and S.~D.~Odintsov, Phys. Rev. D {\bf{68}}, 123512
(2003); P.~S.~Apostolopoulos, N.~Brouzakis, E.~N.~Saridakis and
N.~Tetradis, Phys. Rev. D {\bf{72}}, 044013 (2005); S.Nojiri and
S.~D.~Odintsov, Int. J. Geom. Meth. Mod. Phys. {\bf{4}}, 115
(2007);
  M.~R.~Setare and E.~N.~Saridakis,
  Phys.\ Lett.\  B {\bf 670}, 1 (2008).


\bibitem{cosmo}  P. J. Peebles and B. Ratra,  Rev. Mod. Phys. {\bf{ 75}},
559 (2003);  J. Kratochvil, A. Linde, E. V. Linder and M.
Shmakova, JCAP  {\bf{0407}} 001 (2004); F.~K.~Diakonos and
E.~N.~Saridakis, [arXiv:0708.3143 [hep-th]].

\bibitem{quint} R. R. Caldwell, R. Dave and P. J. Steinhardt, Phys. Rev. Lett. {\bf{80}},
1582 (1998);  M. S. Turner and M. White, Phys. Rev. D {\bf{56}},
4439 (1997); T. Chiba, Phys. Rev. D {\bf{60}}, 083508 (1999);
Z.~K.~Guo, N.~Ohta and Y.~Z.~Zhang, Phys.\ Rev.\  D {\bf 72},
023504 (2005); Z.~K.~Guo, N.~Ohta and Y.~Z.~Zhang, Mod.\ Phys.\
Lett.\  A {\bf 22}, 883 (2007).

\bibitem{phant} R. R. Caldwell, Phys.
Lett. B {\bf{545}}, 23 (2002); S. Nojiri and S. D. Odintsov, Phys.
Lett.  B {\bf 562}, 147 (2003); S. Nojiri and S. D. Odintsov,
Phys. Rev. D  {\bf{72}}, 023003 (2005);  H. Garcia-Compean, G.
Garcia-Jimenez,  O. Obregon, and C. Ramirez,  JCAP 0807, 016
(2008); M. Jamil, M. Ahmad Rashid, and  A. Qadir, [arXiv:0808.1152
[astro-ph]].

\bibitem{phantBigRip}
R.~R.~Caldwell, M.~Kamionkowski and N.~N.~Weinberg, Phys. Rev.
Lett. {\bf 91}, 071301 (2003).

\bibitem{quintom}Z. K. Guo, {\it{et al.}},
Phys. Lett. B {\bf 608}, 177 (2005);
 J.-Q. Xia, B. Feng
and X. Zhang, Mod. Phys. Lett. A {\bf 20}, 2409 (2005); M.-Z Li ,
B. Feng, X.-M Zhang, JCAP, 0512, 002, (2005); B. Feng, M. Li,
Y.-S. Piao and X. Zhang, Phys. Lett. B {\bf 634}, 101 (2006); M.
R. Setare, Phys. Lett. B {\bf 641}, 130 (2006); W. Zhao and Y.
Zhang, Phys. Rev. D {\bf73}, 123509, (2006) ; G.-B. Zhao, J.-Q.
Xia, B. Feng and X. Zhang, Int. J. Mod. Phys. D {\bf16}, 1229
(2007);
 M. R.
Setare, J. Sadeghi, and A. R. Amani, Phys. Lett. B {\bf 660}, 299
(2008); J. Sadeghi, M. R. Setare , A. Banijamali and F. Milani,
Phys. Lett. B {\bf 662}, 92 (2008); M. R. Setare and E. N.
Saridakis, Phys. Lett. B {\bf 668}, 177 (2008); M. R. Setare and
E. N. Saridakis, [arXiv:0807.3807 [hep-th]]; M. R. Setare and E.
N. Saridakis, JCAP {\bf 09}, 026 (2008).



\bibitem{quinteflat}
K. Griest, \prd {\bf 66}, 123501 (2002); S. Bludman, \prd {\bf
69}, 122002 (2004); R.R. Caldwell and E.V. Linder, \prl {\bf 95},
141301 (2005); E.V. Linder, \prd {\bf 73}, 063010 (2006); S.
Chongchitnan and G. Efstathiou, \prd {\bf 76}, 043508 (2007).



\bibitem{Kujat}
  J.~Kujat, R.~J.~Scherrer and A.~A.~Sen,
  Phys.\ Rev.\  D {\bf 74}, 083501 (2006)
  [arXiv:astro-ph/0606735].


\bibitem{Huang}
  Q.~G.~Huang,
  Phys.\ Rev.\  D {\bf 77}, 103518 (2008);
  E.~N.~Saridakis,
  [arXiv:0811.1333 [hep-th]].


\bibitem{Scherrer:2007pu}
  R.~J.~Scherrer and A.~A.~Sen,
  Phys.\ Rev.\  D {\bf 77}, 083515 (2008).

\bibitem{Scherphan}
  R.~J.~Scherrer and A.~A.~Sen,
  Phys.\ Rev.\  D {\bf 78}, 067303 (2008).

\bibitem{Copeland}
E.~J.~Copeland, A.~R.~Liddle and D.~Wands, Phys.\ Rev.\ D {\bf
57}, 4686 (1998).

\bibitem{Crittenden}
  R.~Crittenden, E.~Majerotto and F.~Piazza,
  Phys.\ Rev.\ Lett.\  {\bf 98}, 251301 (2007).

\bibitem{Lindp}
M. Chevallier and D. Polarski, Int. J. Mod. Phys. D {\bf 10}, 213
(2001); E.V. Linder, \prl {\bf 90}, 091301 (2003).

\bibitem{negpot}
B. Ratra and P.J.E. Peebles, \prd {\bf 37}, 3406 (1988); R.R.
Caldwell, R. Dave, and P.J. Steinhardt, \prl {\bf 80}, 1582
(1998).



\end{thebibliography}
\end{document}